\documentclass[aps,prl,twocolumn,showpacs,groupaddress]{revtex4-1}
\usepackage{amsmath}
\usepackage{amsfonts}
\usepackage{multirow}
\usepackage{float}
\usepackage{graphicx}
\usepackage{nicefrac}
\usepackage{subfigure}
\usepackage{xcolor}
\usepackage{bm}
\usepackage{tikz-cd}
\usepackage[breaklinks=true,pdfborder={0 0 0}]{hyperref}
\hypersetup{colorlinks=true,citecolor=red}

\newcommand{\diff}{\mathrm{d}}

\newcommand{\via}{\emph{via }}

\newcommand{\rrr}[1]{{\color{black}#1}}
\newcommand{\bbb}[1]{{\color{black}#1}}
\begin{document}
\title{Casimir Self-Interaction Energy Density of Quantum Electrodynamic Fields}
\author{Alexandre Tkatchenko}
\author{Dmitry V. Fedorov}
\affiliation{Department of Physics and Materials Science, University of Luxembourg, L-1511 Luxembourg City, Luxembourg}
\begin{abstract}
Quantum electrodynamic fields possess fluctuations corresponding to transient particle/antiparticle dipoles, which can be characterized by a non-vanishing polarizability density. Here, we extend a recently proposed quantum scaling law to describe the volumetric and radial polarizability density of a quantum field corresponding to electrons and positrons and derive the Casimir self-interaction energy (SIE) density of the field, $\bar{E}_{\rm{SIE}}$, in terms of the fine-structure constant. The proposed model obeys the cosmological equation of state $w=-1$ and the magnitude of the calculated $\bar{E}_{\rm{SIE}}$ lies in between the two recent measurements of the cosmological constant $\Lambda$ obtained by the Planck Mission and the Hubble Space Telescope. 
\end{abstract}
%
\maketitle

Quantum fields possess transient particle/antiparticle ($p^-/p^+$) pair fluctuations or ``virtual excitations''~\cite{Milonni-book,Zee-book}. 
For electrodynamic fields, such virtual excitations correspond to formation of minute and short-lived dipoles, indicating that the field should possess a finite \emph{intrinsic} polarizability density. 
In this work, we propose a model for the polarizability density of quantum electrodynamic fields based on an extension of a recently derived quantum-mechanical scaling law between polarizability and equilibrium (van-der-Waals) radius~\cite{DimaPRL,alphaJPCL} to $p^-/p^+$ fluctuations. We then calculate the finite Casimir self-interaction energy density of such a quantum field and find that its equation of state 
matches the one of cosmological dark energy, $w=-1$, whereas the magnitude of the obtained
energy density agrees with the most recent measurements of the cosmological constant $\Lambda$.  

Linear polarization of a quantum field is described by a polarizability density-density tensor (PDDT) $\bm{\alpha}_{\rm F}({\bf r}, {\bf r'}, t, t') = \delta \bm{\mathcal{P}}({\bf r}, t) / \delta \bm{\mathcal{E}}({\bf r'}, t')$ -- a functional derivative of polarization field $\bm{\mathcal{P}}({\bf r},t)$ with respect to the electric field $\bm{\mathcal{E}}({\bf r'},t')$. 
\bbb{Since time is homogeneous, $\bm{\alpha}_{\rm F}({\bf r}, {\bf r'}, t, t') = \bm{\alpha}_{\rm F}({\bf r}, {\bf r'}, t-t')$, which allows switching to the frequency-dependent quantity $\bm{\alpha}_{\rm F}({\bf r}, {\bf r'}, \omega)$.}
Upon integrating $\bm{\alpha}_{\rm F}({\bf r}, {\bf r'}, t, t')$ \bbb{or $\bm{\alpha}_{\rm F}({\bf r}, {\bf r'}, \omega)$} over the spatial variable ${\bf r'}$, tracing over tensor components, and taking the infinite-time limit $|t-t'| \to \infty$ \bbb{or the zero-frequency limit $\omega \to 0$} yields a homogeneous static polarizability density $\alpha_{\rm F}({\bf r})$. 
In addition, the following argument suggests that an integral over ${\bf r'}$ leads to a finite value of $\alpha_{\rm F}({\bf r})$. 
\bbb{The production of particle-antiparticle pairs} emerging from the vacuum upon applying \bbb{an electric field \rrr{(Schwinger effect)} is} an established physical fact.
For sufficiently long timescales, the polarizability of \bbb{a created} $p^-/p^+$ pair is essentially infinite since the particle and its antiparticle yield an increasingly growing dipole.
\bbb{In contrast, the polarizability \bbb{of bound $p^-/p^+$ pairs is finite.}}
\rrr{Reducing} the applied field to zero does not eliminate zero-point fluctuations, hence a quantum field should possess intrinsic polarizability density even in the absence of any external field. The observable (di)electric permittivity of vacuum/free space,
$\varepsilon_0$ (with units of polarizability density), has a finite value, 
yielding further support for our proposal.
Obviously, the effects of vacuum polarization are experimentally \bbb{measured by applying} external fields~\cite{ScienceVacPol,Moskalenko2019,BeneaChelmus2019}, hence the possibility of observing intrinsic \bbb{(zero-field)} vacuum polarization is \bbb{a contentious question} at this moment. We are here ultimately concerned with measurable quantities, such as energy density and the cosmological constant. Therefore, we use a finite vacuum polarizability density as a working hypothesis and explore its consequences on experimental observables. In what follows, the polarizability unit will be normalized by $4\pi\varepsilon_0\,$, to measure the polarizability 
$\alpha_{\rm F}$ in terms of the polarizability volume~\cite{Atkins&Friedman_book}. Consequently, the two-point polarizability $\alpha_{\rm F}({\bf r}, {\bf r'})$ has units of 
inverse volume and the polarizability density
$\alpha_{\rm F}({\bf r})$ is unitless. The fine-structure constant (FSC) is denoted by $\alpha_{\rm fsc}\,$.

All polarizable entities experience mutual dispersion forces, which are either of van der Waals (vdW) or Casimir \bbb{type}~\cite{London,Casimir1948,Stone-book,VDW-book,Buhmann-book_I,Buhmann-book_II}. Both forces are manifestations of the same phenomenon related to polarization propagated \via the corresponding gauge field. The broad relevance of dispersion forces has been recognized for diverse phenomena, ranging from particle physics~\cite{Volodya-PRL}, to atoms~\cite{Babb}, molecules~\cite{Jan-ChemRev}, and condensed matter~\cite{Casimir-RMP,LongRange-RMP}, and even for cohesion in macroscopic objects of cosmological relevance~\cite{Nature-Asteroids,Asteroids}. The difference between vdW and Casimir forces is that the latter name is employed when polarization propagation requires finite time determined by the underlying gauge field (usually, speed of light). 

The exact vdW/Casimir self-interaction energy (SIE) density caused by particle/antiparticle fluctuations of a quantum field can be computed using the adiabatic connection fluctuation-dissipation theorem (see Refs.~\cite{ACFDT1,ACFDT2} for seminal works formulated in terms of the density response function and Refs.~\cite{Jan-ChemRev,Woods-RMP,AR-AT-PRL} for a modern formulation in terms of the polarizability):
\begin{eqnarray}
\label{eqExact}
\bar{E}_{\rm SIE} &=& \frac{\hbar}{2\pi V} \int_0^{\infty} du \int_0^1 d\lambda \int_{V}\int_{V} d\mathbf{r} d\mathbf{r}' \nonumber \\
&\times& \mathrm{Tr}\{[{\bm{\alpha}}_{\lambda}(\mathbf{r},\mathbf{r}',iu)-{\bm{\alpha}}_0(\mathbf{r},\mathbf{r}',iu)] \mathbf{T}(\mathbf{r},\mathbf{r}',iu)\}
\end{eqnarray}
where ${\bm{\alpha}}_0(\mathbf{r},\mathbf{r}',iu)$ is the PDDT corresponding to \emph{bare} $p^-/p^+$ fluctuations, with $u$ as the oscillation frequency. Then, ${\bm{\alpha}}_{\lambda}(\mathbf{r},\mathbf{r}',iu)$ is the PDDT corresponding to $p^-/p^+$ fluctuations interacting with strength $\lambda$ and $\mathbf{T}(\mathbf{r},\mathbf{r}',iu)$ is the dipolar propagator for the bosonic field mediating the interaction (\emph{e.g.}~the retarded dipole potential when considering the electromagnetic field). The integration volume $V$ is infinite, but in practice it is sufficient to integrate over $V$ of the same order as the volume of quantum $p^-/p^+$  
fluctuations, \emph{i.e.}~of the order of Thomson's scattering 
length cubed.
\bbb{We remark that elementary particles are usually considered to be structureless, \emph{i.e.}~they have zero size. However, elementary particles acquire an effective orbital size when they are probed by other particles or fields. For example, the electron orbital has an effective radius of $a_0$ (Bohr's radius) when interacting with a proton. Similarly, the effective electron orbital radius becomes $\alpha_{\rm fsc} a_0$ in an inelastic interaction with a photon \rrr{(Compton scattering)}. Whereas, in an elastic electron--photon interaction \rrr{(Thomson scattering)}, the electron orbital radius becomes $\alpha_{\rm fsc}^2 a_0$. Our work proposes a model for the electron/positron field interacting with the photon field in the low-frequency elastic regime relevant to vdW/Casimir phenomena. In this case, the Thomson's scattering radius ($\alpha_{\rm fsc}^2 a_0$) is the most natural choice for the effective size of the electron (and positron) orbital. Furthermore, Thomson's scattering length defines the onset of measurable effects of quantum fluctuations in quantum field theory~\cite{Milonni-book}.}

The internal polarization dynamics which generate ${\bm{\alpha}}_{\lambda}(\mathbf{r},\mathbf{r}',iu)$ for an interacting quantum field are not known in detail yet. Therefore, we resort to coarse-grained models for
${\bm{\alpha}}_{\lambda}(\mathbf{r},\mathbf{r}',iu)$ taking into
account the main symmetries of a quantum field in its vacuum state, such as its homogeneity and isotropy. In the retarded Casimir regime, the static ($u=0$) full-potential ($\lambda$=1) approximation
${\bm{\alpha}}(\mathbf{r},\mathbf{r}')_{u=0}$ is sufficient. Using either homogeneity or isotropy of the field, in what follows we propose models for volumetric and radial polarizability densities of $p^-/p^+$ fluctuations and use these models for evaluating approximations to Eq.~\eqref{eqExact}.

To simplify the notation, forthcoming derivations will be 
carried out for the electron/positron ($e^-/e^+$) field. Generalization to other quantum fields will be briefly discussed below. 
Due to the homogeneity of zero-point fluctuations, the electron/positron field can be modeled as a condensed overall-neutral homogeneous medium of electrons and positrons with $e^-/e^+$ pairs forming transient dipoles characterized by an isotropic polarizability that follows a quantum-mechanical scaling law~\cite{DimaPRL,alphaJPCL}
\begin{equation}
\label{eqAlphae-e+}
\alpha_{e^-/e^+} = \frac{2}{3} \left(\frac{\alpha_{\rm fsc}^{\nicefrac{1}{3}} \times R_{e^-/e^+}}{a_0}\right)^4 R_{e^-/e^+}^3 \quad ,
\end{equation}
where $a_0$ is Bohr's radius and $R_{e^-/e^+}$ is the equilibrium (vdW) radius of the $e^-/e^+$ pair. The quantum-mechanical scaling of the polarizability $\propto R_{\rm vdW}^7$ and its dependence on the fine-structure constant originates from the dressing of particle/hole excitations by virtual photons of the electromagnetic field, as discussed in Ref.~\cite{alphaJPCL}. The four-dimensional renormalization factor $(\alpha_{\rm fsc}^{\nicefrac{1}{3}} \times R_{e^-/e^+} / {a_0})^4$ corresponds to the polarizability density of the excited electromagnetic field~\cite{L4}. The factor $2/3$ accounts for two contributions: i) both electrons and positrons are polarizable (factor of $2$), ii) the $e^-/e^+$ pair polarizes in a homogeneous field where all polarization directions are degenerate, akin to the jellium model (factor of 1/3)~\cite{Nesbet}. \bbb{As discussed above,}
we set the equilibrium radius $R_{e^-/e^+}$ to the Thomson scattering length, $R_{\rm Th} = \alpha_{\rm fsc}^2 a_0$, at which the electrostatic self-interaction energy of a particle
(${e^2}/{4\pi\varepsilon_0 R_{\rm Th}}$) equals its
rest mass–energy ($m_e c^2$). Finally, the effective volume corresponding to each $e^-/e^+$ pair becomes $V_{e^-/e^+}=2 \times \frac{4 \pi}{3} R_{e^-/e^+}^3\,$, since both electrons and positrons contribute to the polarizability volume.

With a homogeneous model for ${\bm{\alpha}}(\mathbf{r},\mathbf{r}')_{u=0} \equiv \alpha_{e^-/e^+}\,$, the Casimir energy density for two interacting $e^-/e^+$ pairs, as a second-order approximation to
Eq.~\eqref{eqExact}, can be obtained according to Ref.~\cite{Casimir-Polder} as
\begin{equation}
\bar{E}_{\rm SIE}^{(2)} = - \frac{23 \hbar c}{4\pi} \times \frac{\alpha_{e^-/e^+}^2}{(2 R_{e^-/e^+})^7 \times V_{e^-/e^+}} \quad .  
\label{eqC2body}
\end{equation}
Expressing the polarizability in terms of $R_{e^-/e^+}$ yields
\begin{equation}
\bar{E}_{\rm SIE}^{(2)} = - \frac{23 \hbar c}{3072 \pi^2} \times \left(\frac{\alpha_{\rm fsc}^{\nicefrac{1}{3}}}{a_0}\right)^8 \times \left(R_{e^-/e^+}\right)^4  \quad ,  
\label{eqC2body_R4}
\end{equation}
whereas upon substituting the equilibrium vdW radius we obtain
\begin{equation}
\bar{E}_{\rm SIE}^{(2)} = - \frac{23\, \alpha_{\rm fsc}^{\nicefrac{29}{3}}}{3072 \pi^2} \times \frac{\hbar c \alpha_{\rm fsc}}{a_0^4} \bbb{ = - \frac{23\, \alpha_{\rm fsc}^{\nicefrac{29}{3}} E_h}{3072 \pi^2 a_0^3}} \quad ,
\label{eqC2bodynum}
\end{equation}
\bbb{where $E_h = \hbar c \alpha_{\rm fsc} a_0^{-1}$ is the Hartree energy.}
The $\bar{E}_{\rm SIE}^{(2)}$ contribution amounts to a pairwise interaction between $e^-/e^+$ pair fluctuations. Calculating $\bar{E}_{\rm SIE}$ for the field requires summing the pairwise potential over all possible many-body geometries of $e^-/e^+$ fluctuations, which can be approximated as a lattice sum. Hence, the Casimir SIE density of the electron/positron field is given by 
\begin{equation}
\bar{E}_{\rm SIE} = \bar{E}_{\rm SIE}^{(2)} \times N_{\rm eff}\ ,
\label{E_SIE_N_eff}
\end{equation}
where $N_{\rm eff}$ corresponds to a weighted lattice sum of $R^{-7}$ potential over all possible field arrangements of $e^-/e^+$ pairs. 
Performing a more general treatment of the field below, here we determine limiting behaviors of $N_{\rm eff}$ as corresponding to body-centered-cubic (bcc), $N_{\rm eff}^{\rm bcc}=11.05$, and face-centered-cubic (fcc), $N_{\rm eff}^{\rm fcc}=13.36$, lattices calculated from 
\begin{equation}
N_{\rm eff} = \sum\limits_{\mathbf{R}_j \ne \mathbf{0}}
\frac{(2 R_{e^-/e^+})^7}{|\mathbf{R}_j|^7}\ ,
\label{lattice_sum}
\end{equation}
where the sum runs over lattice vectors of the corresponding bcc/fcc structure. The obtained estimates for $N_{\rm eff}$ yield
the following interval of (absolute) values for the Casimir SIE density: $|\bar{E}_{\rm SIE}|= \{1.85 \times 10^{-23}, 2.24 \times 10^{-23}\}$ Ha/Bohr$^3$. 
Remarkably, this energy density range (barring the sign to be discussed below)
seems to agree well with the recent estimates of the vacuum energy density as given by the cosmological constant $\Lambda = \{1.84 \times 10^{-23}, 2.20 \times 10^{-23}\}$ Ha/Bohr$^3$~\cite{Astroconstants}, when taking two different values of the Hubble constant measured by either the Planck Mission (67.66 (km/s)/Mpc)~\cite{Planck} or the Hubble Space Telescope (74.03 (km/s)/Mpc)~\cite{Hubble}.

The uncertainty related to the estimation of $N_{\rm eff}$ can be avoided by making a derivation of the Casimir SIE density based on collective fluctuations of concentric spherical field shells, instead of particle/antiparticle pairs. This model is based on both homogeneity and isotropy of quantum fields. First, we define the radial polarizability density of a thin spherical shell with radius $r$ (centered at $r' = 0$) for the electron/positron field
\begin{equation}
\label{eqRadPol}
\bar{\alpha}_{e^-/e^+}(r) = \frac{4 \pi r^2}{a_0^3} \left(\frac{\alpha_{\rm fsc}^{\nicefrac{1}{3}} \times r}{a_0}\right)^4\, r^3 \quad . 
\end{equation}

The Casimir SIE density for a particle/antiparticle field can now be calculated by summing the SIE over concentric spherical shells from $r=0$ to $r=r_{f}$, where $r_{f}$ is the dressed Thomson scattering length. As shown in
Ref.~\cite{alphaJPCL}, the corresponding (vdW) equilibrium
length $R_{e^-/e^+}$ is renormalized to 
$r_{f} = \alpha_{\rm fsc}^{-\nicefrac{1}{3}} R_{e^-/e^+}\,$, 
which gives us $r_{f} = \alpha_{\rm fsc}^{\nicefrac{5}{3}} a_0\,$. Then, the SIE density for the $e^-/e^+$ field, as an approximation for concentric shells to Eq.~\eqref{eqExact}, 
becomes~\cite{Casimir1948,Casimir-Cavity}
\begin{equation}
\bar{E}_{\rm{SIE}} = - \frac{3 \hbar c}{8 \pi a_0^3} \int_{0}^{r_{f}} \frac{\bar{\alpha}_{e^-/e^+}(r)}{r^4}\, \diff r \quad .
\label{E_SIE}
\end{equation}
Evaluating the integral yields
\begin{equation}
\bar{E}_{\rm{SIE}} = - \frac{\hbar c}{4 a_0^6} 
\left(\frac{\alpha_{\rm fsc}^{\nicefrac{1}{3}}}{a_0}\right)^4
r_f^6 \quad .
\label{E_SIE_r_f}
\end{equation}
Finally, substituting $r_f\,$, we obtain
\begin{equation}
\bar{E}_{\rm{SIE}} = - \frac{\hbar c\, \alpha_{\rm fsc}^{\nicefrac{34}{3}}}{4 a_0^4} = - \frac{1}{4} 
\alpha_{\rm fsc}^{\frac{31}{3}} E_h a_0^{-3} \quad .
\label{eqLambda}
\end{equation}
\rrr{In} atomic units, $|\bar{E}_{\rm{SIE}}| = 2.07 \times 10^{-23}$ Ha/Bohr$^3$. This number lies in between the two recent estimates of the cosmological constant $\Lambda = \{1.84 \times 10^{-23}, 2.20 \times 10^{-23}\}$ Ha/Bohr$^3$~\cite{Astroconstants}. In cosmology, Planck's units are often used, setting $\hbar=1$, $c=1$, and $8 \pi G=1$. Expressing Eq.~\eqref{eqLambda} in Planck's units, we obtain $|\bar{E}_{\rm{SIE}}|=3.31\times10^{-122} \,\, l_{\rm P}^{-2}$, where $l_{\rm P}$ is Planck's length.  

The comparison between Eqs.~\eqref{E_SIE_N_eff} and \eqref{eqLambda}, allows us to derive an effective value for $N_{\rm eff}$, as a weighted lattice interaction sum over all possible field arrangements of particle/antiparticle pairs.
Based on Eqs.~\eqref{eqC2bodynum} and \eqref{eqLambda}, we obtain 
$N_{\rm eff} = \bar{E}_{\rm{SIE}} / \bar{E}^{(2)}_{\rm{SIE}} = \frac{768 \pi^2}{23} \alpha_{\rm fsc}^{\nicefrac{2}{3}} \approx 12.40$. 
This number lies in between the corresponding values calculated for body-centered-cubic and face-centered-cubic lattices above.  

Although $\bar{E}_{\rm{SIE}}$ given by Eqs.~\eqref{eqC2bodynum} or \eqref{eqLambda} has a negative sign, as vdW or Casimir energy should have in a typical geometry, it satisfies the (cosmological) equation of state of an expanding scalar 
field $w = P/\bar{E}_{\rm{SIE}}= -1$, where $P=-dE/dV$ is the pressure. The field has a tendency to expand because of 
\bbb{the linear} dependence of the energy on the volume,
$E_{\rm{SIE}} \propto \bar{E}_{\rm{SIE}} V$,
according to Eq.~\eqref{eqLambda}.
We also remark that matter would experience an effective field with a positive $\bar{E}_{\rm{SIE}}$, because the polarization of matter, as an excitation of the field, is measured with respect to the polarization of the vacuum field which would be described \via ${\bm{\alpha}}_0(\mathbf{r},\mathbf{r}',iu)$ in Eq.~\eqref{eqExact} when computing the vdW/Casimir interactions between two material objects. 

The obtained results rely on \bbb{the applicability of the fully retarded Casimir-Polder energy given by Eq.~\eqref{eqC2body}} for calculating the SIE of a quantum field.
\bbb{Let us evaluate its validity for} distances
comparable to Thomson's scattering length. 
The \bbb{Casimir-Polder} formula is accurate for distances $R$ satisfying
\bbb{the retarded regime,}
$R \gg c/ \omega_{\rm F}$, where $\omega_{\rm F}$ is a characteristic frequency. The value of $\omega_{\rm F}$ can be estimated 
from \bbb{the polarizability of a quantum harmonic oscillator~\cite{Jan-ChemRev}, as}
$\alpha_{\rm F}\, \omega_{\rm F}^2 = e^2 N/m_e$,
where $N$ is the effective particle number contributing to field fluctuations. For the electron/positron field, we obtain that Eq.~(\ref{eqC2body}) is accurate for $R \gg \alpha_{\rm fsc}^{\nicefrac{20}{3}}\, a_0$. This condition is strictly satisfied within the context of our work.   

Both Eqs.~(\ref{eqC2body}) and~(\ref{E_SIE}) employ the dipole approximation to the field response. Some authors argue that the response of a quantum field has only dipolar contributions~\cite{Sernelius} and in fact the PDDT of an arbitrary field (or matter) system can be exactly described by an infinite number of dipolar quantum oscillators~\cite{Milonni-book,Jan-ChemRev}. However, we rely on a coarse-grained representation for the PDDT of a quantum field, which could make multipolar interactions non-negligible~\cite{Jan-ChemRev}. Because of particle/antiparticle symmetry, the quadrupole polarizability vanishes and the first non-vanishing multipolar polarizability beyond dipole is the octupolar one. We can estimate the octupolar contribution to $\bar{E}_{\rm{SIE}}$ by a generalization of Eqs.~\eqref{eqRadPol} and \eqref{E_SIE} to higher multipoles and using the quantum scaling law for multipole polarizabilities derived in Refs.~\cite{DimaPRL,OrnellaPRR}. By doing so, we find that the contribution from a potential octupolar interaction term to $\bar{E}_{\rm{SIE}}$ would be vanishingly small (multiplied by $\alpha_{\rm fsc}^6$) compared to the dipolar one in Eq.~(\ref{E_SIE}). The tiny magnitude of the dipole
polarizability of pair excitations in a quantum field
$\propto \alpha_{\rm fsc}^{\nicefrac{46}{3}}$ in Eq.~\eqref{eqAlphae-e+} provides an explanation of why drastic approximations to the exact vdW/Casimir energy density seem
to yield very accurate results when computing the SIE density of quantum fields in their vacuum state.


\bbb{Going beyond the electron-positron field, we can also calculate the
$\bar{E}_{\rm SIE}$ for muon-antimuon and tau-antitau \rrr{virtual pair} fluctuations.}
\rrr{For heavier leptons possessing the same charge but an arbitrary mass $\mu$, one needs to replace $a_0 = (4\pi\epsilon_0)\hbar^2/m_e e^2$ by $a_{\mu} = (4\pi\epsilon_0)\hbar^2/\mu e^2$ in Eq.~\eqref{eqAlphae-e+}, with a corresponding scaling of the vdW radius: $R_{\mu^-/\mu^+} = (a_{\mu}/a_0)\times R_{e^-/e^+} = (m_e/\mu)\times R_{e^-/e^+}$.~These conditions ensure the correct polarizability-mass scaling for hydrogen-like atoms~\cite{L4}:~$\alpha_{\mu^-/\mu^+} = \alpha_{e^-/e^+} \times (m_e /\mu)^3$.~Since both the polarizability and the related volume of a lepton pair are inversely proportional to the particle mass cubed, the electron-positron, muon-antimuon, and tau-antitau quantum fields possess the same polarizability density $\propto$~$\epsilon_0$, as they should.~Based on Eq.~\eqref{eqC2bodynum}, we obtain $\bar{E}_{\rm SIE}^{(2)} (\mu^-/\mu^+)~/~\bar{E}_{\rm SIE}^{(2)} (e^-/e^+) = (a_0/a_{\mu})^4 = (\mu/m_e)^4$.
However, the total Casimir SIE density given by Eq.~(6) is a product of the pairwise SIE density and the weighted lattice sum $N_{\rm eff}\,$. This effective number of pairwise interacting virtual pairs depends on the lifetime of constituent particles.
Each virtual particle possesses a finite lifetime $\tau_{\mu} = \tau_e \times (m_e/\mu)$, as follows from Heisenberg's uncertainty principle. For two interacting species, four virtual particles are present, which requires the renormalization $N_{\rm eff} (\mu^-/\mu^+)~/~N_{\rm eff} (e^-/e^+) = (m_e/\mu)^4$. This scaling compensates the factor $(\mu/m_e)^4$ in the ratio $\bar{E}_{\rm SIE}^{(2)} (\mu^-/\mu^+)~/~\bar{E}_{\rm SIE}^{(2)} (e^-/e^+)$ resulting in the same Casimir SIE density for electron-positron, muon-antimuon, and tau-antitau fields. Our arguments based on the uncertainty principle are supported by the effective Lagrangian approach in quantum electrodynamics (QED), where each fermionic propagator is inversely proportional to the particle mass and the contributions of Feynman diagrams for (two-)photon exchange between two virtual pairs scale with $\mu^{-4}$~\cite{Greiner-book}. In particular, the first non-linear term in the effective Heisenberg-Euler Lagrangian~\cite{Greiner-book,Schwinger,Dittrich} describing interactions between quadratic field fluctuations scales as $\mu^{-4}$. This fourth-order QED process is closely connected to our Casimir model~\cite{Reza_PRR,CraigThirunamachandran-book,Salam-book}, for both Eq.~\eqref{E_SIE_N_eff} and Eq.~\eqref{eqLambda}. Thus, we conclude that electrodynamic fields corresponding to different leptons possess the same Casimir SIE density.}

The extension of our work to fields other than the \bbb{ones of charged leptons} 
would require accounting for propagators other than the electromagnetic
one in Eq.~\eqref{eqExact} and for the increase of the FSC at high energies~\cite{Fritzsch2002}. However, the conceptual idea underlying the presented calculations should be equally applicable to arbitrary fields. In the current model, the energy density of a given field depends on two partially compensating factors determined by the characteristic length of quantum fluctuations $R_{\rm Th}$: (i) the polarizability density that scales $\propto$ $R_{\rm Th}^7$, (ii) the (pairwise) Casimir interaction that scales $\propto$ $R_{\rm Th}^{-7}$. Ultimately, a scale-invariant value of SIE density equal to $\Lambda$ for any quantum field would be the most elegant result and such outcome could arise from a subtle compensation between the polarizability density and the Casimir interaction potential. 

To embed our work in the context of other models for the cosmological constant~\cite{CC-RMP}, we note that there is a solid body of literature attempting to connect the cosmological and gravitational constants to zero-point fluctuations of quantum fields. These attempts go back to seminal phenomenological proposals by Dicke~\cite{Dicke}, Zel'dovich~\cite{Zeldovich} and Sakharov~\cite{Sakharov}. However, calculations based on the direct summation of zero-point energies of quantum fields overestimate the observed cosmological constant by 40 to 120 orders of magnitude~\cite{CC-RMP}. Interestingly, if instead of the quantum scaling law for polarizability in Eq.~\eqref{eqAlphae-e+} we use the classical scaling $\alpha_{e-/e+} \propto R_{e-/e+}^3$, the resultant $\bar{E}_{\rm SIE}$ would overestimate $\Lambda$ 
by 39 orders of magnitude, similar to the simplistic direct sum over zero-point energies~\cite{Hawking}. Modified approaches to calculation of zero-point energies have been attempted in the past, for example by Puthoff~\cite{Puthoff1,Puthoff2}, and more recently by Leonhardt, who proposed to apply Lifshitz theory with renormalization and obtained a more reasonable estimate for the energy density of the electromagnetic vacuum field~\cite{Leonhardt1,Leonhardt2}.

Instead, the key conceptual advance in our work is to propose that one can define non-interacting particle/antiparticle fluctuations, with the PDDT given by
${\bm{\alpha}}_0(\mathbf{r},\mathbf{r}',iu)$. These fluctuations then interact through the electromagnetic field, yielding ${\bm{\alpha}}_{\lambda=1}(\mathbf{r},\mathbf{r}',iu)$, whose coarse-grained representation is given by the rather minute homogeneous polarizability in Eq.~\eqref{eqAlphae-e+} or the radial shell polarizability density in Eq.~\eqref{eqRadPol}. The Casimir interaction propagated by the respective gauge field then yields a finite SIE density for any fermionic quantum field. It is pertinent to draw a parallel between our model
and the calculation of many-electron correlation energies
in molecules and materials based on Eq.~\eqref{eqExact}~\cite{Dobson,Kresse,Xinguo-RPA}. 
For interacting electrons in matter, the definition of ${\bm{\alpha}}_0(\mathbf{r},\mathbf{r}',iu)$ has a high degree of arbitrariness, as it can be constructed from Hartree-Fock, Kohn-Sham, Wannier orbitals, or alternatively using quantum harmonic oscillators to model the response of valence electrons~\cite{Jan-ChemRev,Woods-RMP}. 
Hence, ${\bm{\alpha}}_0(\mathbf{r},\mathbf{r}',iu)$ or its coarse-grained representations are unobservable. In contrast, the interacting macroscopic polarizability tensor \bbb{$\int_{\bf r} \int_{{\bf r}'} {\bm{\alpha}}_{\lambda=1}(\mathbf{r},\mathbf{r}',iu)\, d{{\bf r}'} d{\bf r}$} is observable in practice. We are not aware of
experimental measurements of the \textit{intrinsic} polarizability density of the electron/positron field, however our work makes a verifiable prediction that quantum fields possess a non-vanishing (albeit quite small) polarizability density.       

In summary, we have proposed a model for quantum electrodynamic fields endowed with a finite polarizability density arising from zero-point fluctuations of particle/antiparticle pairs. We then calculated the Casimir self-interaction energy density of such quantum fields (owing to propagation of fluctuations by the electromagnetic field) and obtained the expected dark-energy equation of state and excellent numerical agreement with the measured value of the cosmological constant $\Lambda$. 
Obviously, our work leaves many open questions concerning the polarizability density
of more general quantum fields and its contribution to the vacuum self-interaction energy density. \rrr{The preliminary connections identified in this work between the Casimir approach and the effective Lagrangian formalism in QED suggest a possibility to calculate the self-interaction energy density directly from first principles of quantum field theory.}

The authors acknowledge financial support from the European Research Council
via the ERC Consolidator Grant \lq\lq BeStMo(GA 725291)\rq\rq\ and the Luxembourg National Research Fund via the FNR CORE Jr project \lq\lq PINTA(C17/MS/11686718)\rq\rq. We thank Matteo Gori (Uni.lu), Matthieu Sarkis (Uni.lu), Jorge Charry (Uni.lu), Alejandro Rodriguez (Princeton), Lilia Woods (USF), Andrey Moskalenko (KAIST), and John F. Dobson (Griffith) for valuable discussions. We also thank the Anonymous Referee B for the insightful feedback that helped us improve our presentation. A.T. dedicates this paper to a decade of friendship and collaboration with Klaus-Robert M\"uller (TU Berlin).

\bibliographystyle{apsrev4-1}
\bibliography{literature}
\end{document}